\documentclass{PoS}

\title{Quark tensor and axial charges within the Schwinger-Dyson formalism}

\ShortTitle{Quark tensor and axial charges within the Schwinger-Dyson formalism}

\author{\speaker{Nodoka~Yamanaka}, %\thanks{A footnote may follow.},
 Takahiro~M.~Doi, Shotaro~Imai, Hideo~Suganuma\\
        Department of Physics, Graduate School of Science,
  Kyoto University, \\
  Kitashirakawa-oiwake, Sakyo, Kyoto 606-8502, Japan\\
        E-mail: \email{yamanaka@ruby.scphys.kyoto-u.ac.jp}}

\abstract{
We calculate the tensor and axial charges of the quark in the Schwinger-Dyson formalism of Landau gauge QCD.
It is found that the dressed tensor and isovector axial charges of the quark are suppressed against the bare quark contribution, and the result agrees qualitatively with the experimental data.
We show that this is due to the superposition of the spin flip of the quark arising from the successive emission of gluons which dress the vertex.
For the isoscalar quark axial charge, we have analyzed the Schwinger-Dyson equation by including the leading unquenching quark-loop effect.
It is found that the suppression is more significant, due to the axial anomaly effect.
}

\FullConference{XV International Conference on Hadron Spectroscopy-Hadron 2013\\
		4-8 November 2013\\
		Nara, Japan }

\begin{document}

\section{\label{sec:intro}Introduction}

The analysis of the nucleon parton structure plays an essential role in clarifying the dynamics of the quantum chromodynamics \cite{review}.
The spin-dependent quark distributions of the nucleon in the leading twist are given by 
the helicity and the transversity distributions functions of the quark, 
and their first moments are called the axial charge ($\Delta q$) and the tensor charge ($\delta q$), respectively.
In the nonrelativistic constituent quark model, which considers three massive quarks in the nucleon, the axial and tensor charges of the quark in the proton become their spin, so that we have $\Delta u = \delta u = \frac{4}{3}$ for the $u$ quark, and $\Delta d = \delta d = -\frac{1}{3}$ for the $d$ quark.
The quark axial and tensor charges in the nucleon are chiral-even and -odd quantities, respectively, and their difference probes how relativistic the polarized quarks are.
Then, their study is important in understanding the structure of the nucleon.
We also note that the quark tensor charge relates the quark electric dipole moment (EDM) to the nucleon EDM, an observable sensitive to the CP violation of the elementary interactions, and is thus an important quantity in the search of new physics beyond the standard model \cite{edm}.

The experimental study of these charges for the proton, however, gives smaller results compared to the naive quark model prediction, namely \cite{recentcompass,ucn,tensorextraction}
\begin{eqnarray}
\Delta \Sigma &=& 0.32 \pm 0.03 \pm 0.03 
\, ,
\\
g_A &=& -1.27590 \pm 0.00239\, ^{+0.00331}_{-0.00377} 
\, ,
\label{eq:g_Aexp}
\\
\delta u &=& 0.860 \pm 0.248 \, , 
\label{eq:deltauexp}
\\
\delta d &=& -0.119 \pm 0.060 \, ,
\label{eq:deltadexp}
\end{eqnarray}
where $\Delta \Sigma \approx \Delta u +\Delta d$ and $g_A =\delta u - \delta d$.
Here, the quark tensor charges were renormalized at $\mu = 2$ GeV.
Also the lattice QCD studies of the quark axial and tensor charges give smaller results than the naive quark model prediction \cite{latticereview}, in qualitative agreement with the experimental data.
We should therefore try to clarify the source of this suppression with some nonperturbative method.

As a powerful nonperturbative way to investigate the dynamics of the quantum field theory and in particular the low energy QCD, we have the Schwinger-Dyson (SD) formalism, and many studies such as the dynamical quark mass, the meson masses, the form factors, etc, have been done so far \cite{higashijima,robertsreview,tensorsde,axialsde}.
In this paper, we will try to clarify the effect of the gluon vertex dressing and analyze the source of the deviation of the quark charges.
The effect in question, the vertex gluon dressing, is well within the applicability of the SD formalism.

\section{\label{sec:setup}Basics of the SD Formalism}

In this work, we consider the SD formalism of the Landau gauge QCD with the rainbow-ladder approximation.
Here, the SD formalism includes the infinite order of the strong coupling, i.e. nonperturbative effects, and the quark-gluon vertex is also renormalization group (RG)-improved at the one-loop which gives the replacement $\frac{g_s^2}{4\pi} Z_g (q^2) \gamma^\mu \times \Gamma^\nu (q , k ) \rightarrow \alpha_s (q^2) \gamma^\mu \times \gamma^\nu$ where $Z_g (q^2)$ is the gluon dressing function, and $\Gamma^\nu (q,k) $ is the dressed quark-gluon vertex.
We use the RG-improved strong coupling $\alpha_s (p^2)$ with infrared (IR) regularization \`{a} la Higashijima \cite{higashijima}
\begin{eqnarray}
\alpha_s (p^2) =
\left\{
\begin{array}{ll}
\frac{24\pi}{11N_c - 2N_f} & (p<p_{\rm IR}) \cr
\frac{12\pi}{11N_c - 2N_f} \frac{1}{\ln (p^2 / \Lambda_{\rm QCD}^2)} & (p\geq p_{\rm IR}) \cr
\end{array}
\right.
\, .
\end{eqnarray}
Here we take $N_c =N_f = 3 $ and $p_{\rm IR}$ satisfying $\ln (p_{\rm IR}^2 / \Lambda_{\rm QCD}^2)= \frac{1}{2}$.
We use the QCD scale parameter $\Lambda_{\rm QCD}$ = 900 MeV.
This large scale parameter is taken to reproduce the chiral quantities in this framework \cite{tensorsde}.
With this setup, the chiral condensate (renormalized at $\mu = 2$ GeV) is given by $\langle \bar q q \rangle = -(238$ MeV)$^3$ and $f_\pi \simeq $ 70 MeV.

\begin{figure}[htb]
\begin{center}
\includegraphics[width=10cm]{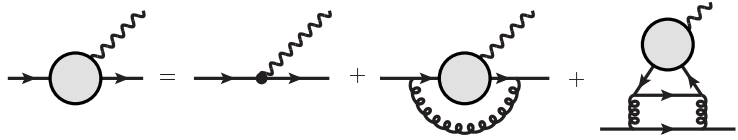}
\caption{\label{fig:SDE}
The Schwinger-Dyson equation for the quark axial and tensor charges expressed diagrammatically.
The grey blobs represent the dynamical charges, and the black dot the bare charge.
Note that the last unquenching diagram does not contribute to the tensor and isovector axial charges.
}
\end{center}
\end{figure}

The Schwinger-Dyson equation (SDE) for the axial and tensor charges are depicted in Fig. \ref{fig:SDE}.
The calculated dynamical charges give the contribution of the single quark to the corresponding nucleon charges.
For the exact expressions of SDE and details, we refer the reader to Refs. \cite{tensorsde,axialsde}.

\section{\label{sec:analysis}Analysis}

\begin{figure}[b]
\begin{center}
\includegraphics[width=5.8cm,angle=-90]{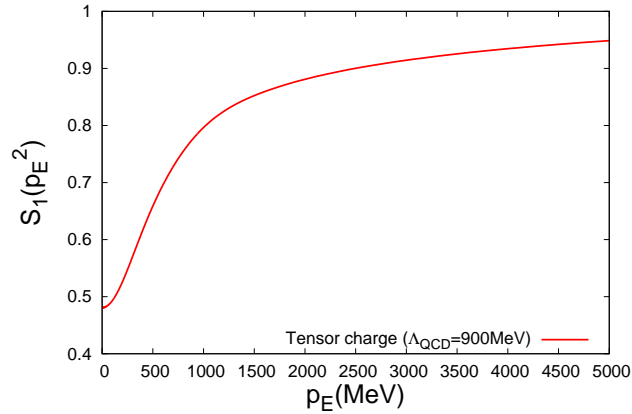}
\caption{\label{fig:tensorsde} 
The dressing function of the dynamical quark tensor charge $S_1 (p_E) \sigma^{\mu \nu}$ (not renormalized) obtained after solving the SDE.
The horizontal axis denotes the Euclidean momentum.
}
\end{center}
\end{figure}

\begin{figure}[htb]
\begin{center}
\includegraphics[width=5.8cm,angle=-90]{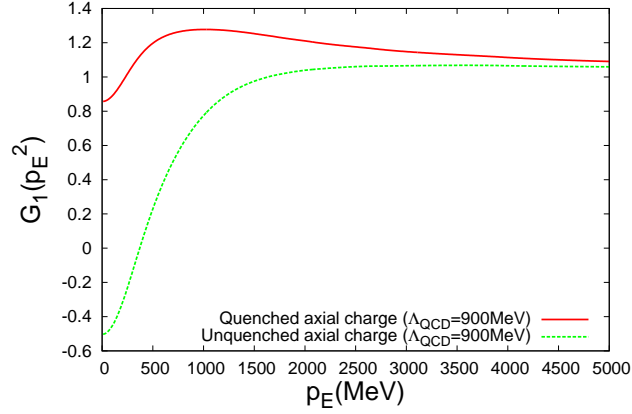}
\caption{\label{fig:axialsde}
The dressing function of the dynamical quark axial charge $G_1 (p_E) \gamma^\mu \gamma_5$ obtained after solving the SDE.
The horizontal axis denotes the Euclidean momentum.
}
\end{center}
\end{figure}

The solutions of the quark tensor and axial SDE are shown in Figs. \ref{fig:tensorsde} and \ref{fig:axialsde}, respectively.
If we associate the dressed dynamical quark with the constituent quark, our result can be combined with the nonrelativistic constituent quark model prediction.
For the quark tensor charge, we have
\begin{eqnarray}
\delta u &=& \frac{4}{3} S_1 (0)_\mu \simeq 0.8\, ,
\nonumber\\
\delta d &=& -\frac{1}{3} S_1 (0)_\mu \simeq -0.2 \, ,
\, ,
\label{eq:(15)}
\end{eqnarray}
where the above tensor charges are renormalized at $\mu = 2 $ GeV.
In the above derivation, it is assumed that the nucleon is composed of three constituent valence quarks with negligible spin-dependent many-body interactions.
The suppression of the tensor charge agrees qualitatively with the results obtained from the extraction from experimental data (\ref{eq:deltauexp}) and (\ref{eq:deltadexp}) \cite{tensorsde}.

For the quark axial charge, we have
\begin{eqnarray}
g_A 
&=&
\frac{5}{3} G_1 (0) \simeq
 1.43 \, ,
\label{eq:g_A}
\\
\Delta \Sigma 
&=&
G_1 (0)_\mu \simeq
 -0.47 \, ,
 \label{eq:totspin}
\end{eqnarray}
where we have again combined the solution of the SDE with the constituent quark model argument.
For the isovector quark axial charge $g_A$, the dynamical axial charge is suppressed compared with the bare one.
The result of Eq. (\ref{eq:g_A}) is in qualitative agreement with the experimental value (\ref{eq:g_Aexp}).

\begin{figure}[htb]
\begin{center}
\includegraphics[width=11.3cm]{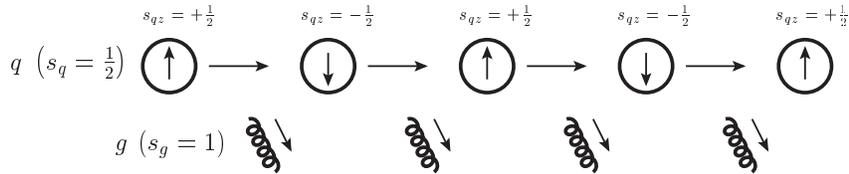}
\caption{\label{fig:quark_spin}
The schematic picture of the quark spin flip with the gluon emission/absorption.
}
\end{center}
\end{figure}

The suppression of the quark tensor and isovector axial charges can be explained by the spin flip of the quark after each emission or absorption of the gluon (see Fig. \ref{fig:quark_spin}).
This mechanism is consistent with the angular momentum conservation since the quarks and gluons have spin one half and one, respectively.
By outputting the tensor and axial charges after each iteration of the SDE, we have confirmed that the result converges by oscillating.
This shows that the reversal of the quark spin is preferred in the gluon emission/absorption, and is thus consistent with our analysis.

As shown in Eq. (\ref{eq:totspin}), the dynamical isoscalar quark axial charge $\Delta \Sigma$ is also suppressed against the bare one, and we see that $\Delta \Sigma$ is much more suppressed than $g_A$ , well below zero.
The unquenching diagram of Fig. \ref{fig:SDE} actually gives the axial anomaly contribution, and this result suggests that the axial anomaly has a significant effect in the suppression of the isoscalar quark axial charge.

\section{\label{sec:summary}Summary}

In this paper, we have calculated the quark axial and tensor charges in the SD formalism of the Landau gauge QCD with RG-improved rainbow-ladder truncation.
For the quark tensor and isovector axial charges, our result suggests that the gluon dressing of the vertex suppresses the charges, and is in qualitative agreement with the experimental data.
The isoscalar quark axial charge receives a larger suppression due to the axial anomaly through the unquenching quark-loop.
This unquenching effect may be largely overestimated due to the large uncertainty in treating the IR region.
For, the axial anomaly also contributes to the many-body effect via the exchange interaction.
To study the problem of the proton spin quantitatively, we must therefore evaluate the many-body effect together with the discussion of this paper.

\acknowledgments{
This work is in part supported by the Grant for Scientific Research [Priority Areas ``New Hadrons'' (E01:21105006), (C) No.23540306] from the Ministry of Education, Culture, Science and Technology (MEXT) of Japan.
}

\end{document}